\documentclass[useAMS,usenatbib,usegraphicx]{mn2e}
\usepackage{longtable}
\usepackage{times}
\usepackage{mathptmx}
\usepackage[usenames,dvipsnames]{color}

\newcommand{\um}[1]{\,\mathrm{#1}}
\newcommand{\mic}[1]{\,\umu\mathrm{#1}}

\title[SCORPIO: A deep radio survey from the stellar life-cycle]
{SCORPIO: \\A deep survey of Radio Emission from the stellar life-cycle}
\author[G. Umana et al.]
  {G.~Umana$^1$\thanks{email: Grazia.Umana@oact.inaf.it},
  C.,~Trigilio$^1$,  T.M.O.~Franzen$^2$,  R.P.~Norris,$^2$, P.~Leto$^1$, A.~Ingallinera$^{1}$,
 \newauthor  C.S.~Buemi$^1$, C.~Agliozzo$^{4,5}$, F.~Cavallaro$^{3,1,2}$, L.~Cerrigone$^6$\\
  $^1$INAF- Osservatorio Astrofisico di Catania, Via S. Sofia 78,  95123, Catania, Italy\\
  $^2$CSIRO Astronomy and Space Science, PO Box 76, Epping, NSW 1710, Australia\\
  $^3$Universit\`a di Catania, Dipartimento di Fisica e Astronomia, Via Santa Sofia, 64, 95123 Catania, Italy\\
  $^4$Millennium Institute of Astrophysics, Santiago, Chile\\
  $^5$Universidad Andr\'es Bello, Avda. Republica 252, Santiago, Chile\\
  $^6$ASTRON, the Netherlands Institute for Radioastronomy, PO Box 2, 7990 AA Dwingeloo,The Netherlands}
 
\begin{document} 
\date{\textbf{}}

\pagerange{\pageref{firstpage}--\pageref{lastpage}} \pubyear{2015}

\maketitle
\label{firstpage}

\begin{abstract}

Radio emission has been detected in a broad variety of stellar objects from all stages of stellar evolution. However, most of our knowledge originates from targeted observations of small samples, which are strongly biased to sources which are peculiar at other wavelengths.  
 In order to tackle this problem we have conducted a deep $1.4\um{GHz}$ survey by using the Australian Telescope Compact Array (ATCA), following the same observing setup as that used for the Australia Telescope Large Area Survey (ATLAS) project, this time choosing a region more appropriate for stellar work. In this paper, the SCORPIO project  is presented as well as results from the pilot experiment. 
The achieved rms is about $30 \mic{Jy}$ and the angular resolution $\sim 10$ arcsec.
 About six hundred of point-like sources have been extracted just from the pilot field. A very small percentage of them are classified in SIMBAD or the NASA/IPAC Extragalactic Database (NED). About $80 \%$ of the extracted sources are reported in one of the inspected catalogues and $50\%$ of them appears to belong to a reddened
 stellar/galactic population. The evaluation of extragalactic contaminants is very difficult without further investigations.
 Interesting results have   been  obtained for extended radio sources that fall in the SCORPIO field. Many bubble-like structures have been found, some of which are classified at other wavelengths. However, for all of these sources, our project has provided us with images of unprecedented sensitivity and angular resolution at $2.1\um{GHz}$.

\end{abstract}

\begin{keywords}
Galaxy: stellar content -- Radio continuum: stars -- radio continuum: ISM -- stars: evolution -- stars: formation -- techniques: interferometric.
\end{keywords}

\section{Introduction}
In the last few years, large radio surveys such as the NRAO VLA Sky Survey (NVSS) \citep{Condon98}, the Faint Images of the Radio Sky at Twenty-centimeters (FIRST) \citep{Helfand99} and ATLAS \citep{Norris06, Hales14} have revealed different populations of radio emitting objects. However, very few stars have been found and nearly all known radio stars have been detected by targeted  observations directed at small samples of stars thought likely to be radio emitters. The radio detection of different star types, covering different stages of stellar evolution \citep{Gudel02}, strongly suffers from  limited sensitivity  and from selection bias as they have been obtained by targeted observations  aimed  to  address a specific problem related to a particular aspect of radio emission.

This approach has  proven to be quite productive but it is biased against discovering unknown, unexpected, or intrinsically rare objects, preventing a good knowledge of radio stars at the sub-mJy level. In consequence, at the moment, it is quite difficult to provide a   trustworthy  forecast  on the real potential of the Square Kilometer Array (SKA) and its pathfinders in the field of stellar radio astronomy. From an analysis aimed to point out the stellar population in the FIRST survey, \cite{Helfand99} concluded that even at a threshold of $\sim\!1\um{mJy}$ stellar radio emission is quite rare. Similar results were found  by \citet{Kimball09}. These results are not surprising as FIRST and other radio deep surveys  were designed for extragalactic studies and only regions at high Galactic latitude were covered. The space density of stars, unlike that of extragalactic sources, varies with Galactic latitude, being much more concentrated in the disk. To overcome this problem, it is clear that a deep blind survey, carried out in a sky patch well suited for stellar work, is necessary.

\subsection{Stellar radio emission}
The improvement of the observational capabilities have led to the discovery of radio emission in  a broad variety of stellar objects from all stages of stellar evolution. In many cases, radio observations have revealed astrophysical phenomena and stellar activity not detectable by other means. Broadly speaking, the brightest stellar radio  emission appears to be  associated with active stellar phenomena such as flares, related to the presence of a strong and/or variable magnetic field, or mass-loss \citep{seaquist_93, Gudel02}. Much of our knowledge of microwave  emission from radio stars comes from the study of active stars and binary systems as a large fraction of these  have been found to be strong radio sources. Their radio flux density is highly variable  and is very probably driven by the magnetic activity whose manifestations are observed in other spectral regions \citep{drake_89,umana_93,umana_95,White_2004}. The radio flux arises from the interaction between the stellar magnetic field and  mildly relativistic particles, i.e. gyrosynchrotron emission \citep{dulk85}.  The same emission mechanism is at the origin of radio emission from pre-main sequence (PMS) stars and X-ray binaries. Non-thermal radio emission also originates from shocks of colliding winds in massive binaries and from pulsars. There is  growing evidence that radio flares can also occur  as narrow band, rapid, intense and highly polarized (up to $100\%$) radio bursts, that are observed especially at low frequency ($<\!1.5\um{GHz}$). For their  extreme characteristics, such radio flares have generally been interpreted as a result of coherent emission mechanisms. Coherent burst emission has been observed in different classes of stellar objects: RS CVns and Flare stars \citep{Osten2004,slee_08}, Ultra Cool  dwarfs \citep{hallinan_08, route_12} and Chemically Peculiar (CP) stars \citep{trigilio_08, trigilio_11}. All have, as common ingredient, a strong magnetic field, which may be variable,  and a source of energetic particles. The number of stars where coherent emission has been detected is still limited to a few tens, because of the limited sensitivity of the available instruments. Thermal emission (bremsstrahlung emission) is expected from winds associated with Wolf--Rayet (WR) and OB stars, shells surrounding planetary nebulae (PNe) and novae and jets from symbiotic stars and class 0 pre-main-sequence (PMS) stars \citep{White_2000}.

\section{The SCORPIO project}
We have started the
Stellar Continuum Originating from Radio Physics In Ourgalaxy (SCORPIO)
project to carry out, for the first time, a blind 
deep ($30 \mic{Jy}$) 
radio survey in a sky patch well suited for stellar work, using ATCA observations at $1.4\um{GHz}$.

This survey  has the best potential to enlarge the stellar radio emitting population, without suffering from selection criteria based on some peculiar aspects observed in other spectral bands. Moreover, it will provide new insights for a better comprehension of the physics of particular classes of stellar systems and of plasma processes in a wider context. Finally, the chosen frequency (i.e. $1.4\um{GHz}$) constitutes the best choice to answer the key question of how common is coherent radio emission from stellar and sub-stellar systems. As the space distribution of different types of stars in the patch selected for our survey can be assumed to be typical for the Galactic disk population, results from such a survey will provide us with a clear forecast on the potential of SKA and its pathfinders in the field of stellar radio astronomy.

In addition to the scientific outcomes, results from the survey will be of immense value in shaping the  strategy  of the deep ($\sim\!10\mic{Jy}$) surveys already planned with the SKA pathfinders. Among them, the Evolutionary Map of the Universe (EMU; \citealp{Norris_2011}), a deep, almost full sky (75 per cent) survey to be carried out at $1.4\um{GHz}$, is one  of the major programs already approved to be conducted with the Australian SKA Pathfinder (ASKAP). EMU is expected to detect and catalogue about 70 million galaxies, and will also make an atlas of the Galactic Plane to an unprecedented sensitivity and resolution. 

The SCORPIO project will have a profound impact on the Galactic Plane component of the EMU survey. Specifically, it will guide EMU design in identifying  issues arising from the complex continuum structure associated with the Galactic Plane and from the variable sources in the Galactic Plane. Moreover, it will contribute in evaluating the most appropriate method for source finding and extraction for sources embedded in the diffuse emission expected at low Galactic latitude.

\subsection{The selected field}
The selected field needs to satisfy the following  requirements: (i) it should contain a sufficient number of stars, with a good spread in different classes of stellar objects thought to be radio emitters; (ii) it should contain sources already classified as radio emitters to enable verification of the data; (iii) it should have  been already observed in other spectral regions to gather additional information to help with the classification of new, unexpected objects or to allow complete studies for the classified objects.

The selected field is namely $2\times2\um{deg}^2$ region centred at galactic coordinates $l=344.25^\circ$, $b=0.66^\circ$. Because our aim is to get statistical information from the survey,  we choose this sky patch only  on the a priori knowledge of the presence of a significative number of stars in it. Moreover, being  in the direction of the Scorpio constellation, it  offers the possibility to probe different radio star populations at various  distances across the Galaxy. As added value, in the field there is also the Sco OB1 association, whose core is the young  ($\sim\!3\!-\!5\um{Myr}$) stellar cluster NGC 6231 \citep{perry1990} ($D \sim 1.6\um{kpc}$), consisting of 964 stars, where radio continuum  emission from the massive star population has been already detected \citep{Setia2003} and X-ray properties of both OB  and PMS populations have been derived \citep{Sana2006}. There are, in total, eight open clusters of stars and stellar associations inside the SCORPIO field, as shown in Table\,\ref{cluster}. [DBS2003]~176 and [DBS2003]~178 are two clusters of stars detected in the Two-Micron All Sky Survey (2MASS) \citep{dutra2003}. C~1653-405 and C~1658-410 are two young open clusters with young O-type stars. DSH~J1704.3-4204 is a cluster candidate, discovered in the Digitized Sky Survey (DSS) and 2MASS survey \citep{kron2006}, that is situated in a heavily obscured part of the Milky Way; it could correspond to  a region of lower extinction rather than a physical group of star. MCM2005b~86 is a star cluster discovered in the Galactic Legacy Infrared Mid-Plane Survey Extraordinaire (GLIMPSE) \citep{mercer2005}.

\begin{table}
\caption{Stellar Clusters and Associations in the SCORPIO field}
\begin{center}
\begin{tabular}{lrrrr}
\hline
Name      &  RA       & Dec        & $l$ & $b$ \\
               &   (J2000)& (J2000) & (deg)   &  (deg)  \\ 
\hline
NGC 6231                   &  16 54 08.5  &  $-$41 49 36 & 343.4601 &  +01.1866   \\
Sco OB 1                   &  16 53 28.8  &  $-$41 57 00 & 343.2900 &  +01.2000   \\
$[$DBS2003$]$ 176          &  16 59 23.0  &  $-$42 34 23 & 343.4830 &  $-$00.0380   \\
$[$DBS2003$]$ 178          &  17 02 10.1  &  $-$41 46 48 & 344.4260 &  +00.0440   \\
C 1653-405                 &  16 57 00.0  &  $-$40 40 01 & 344.7010 &  +01.4960   \\
C 1658-410                 &  17 02 12.0  &  $-$41 04 12 & 344.9900 &  +00.4800   \\ 
DSH J1704.3-4204           &  17 04 20.1  &  $-$42 04 24 & 344.4394 &  $-$00.4546   \\
MCM2005b 86                &  17 04 40.1  &  $-$41 42 25 & 344.7680 &  $-$00.2810   \\
\hline
\end{tabular}
\end{center}
\label{cluster}
\end{table}

\begin{figure}
\includegraphics[width=7cm]{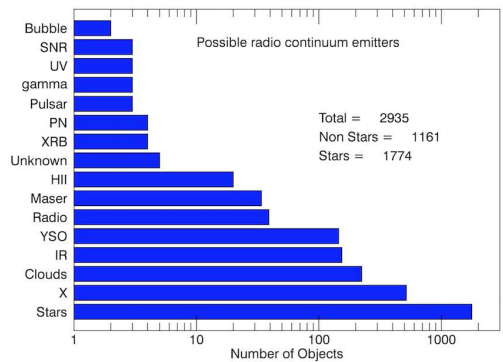}
\caption{Histogram showing the number of each different possible radio emitting class, listed in SIMBAD, falling in the selected field. The labels follow the SIMBAD nomenclature. There are many sources with only general X-ray, UV and IR  classification. Bubbles, SNR, $\gamma$-ray sources, Pulsars, PN, H 	\textsc{ii} regions and High and Low mass X-ray binaries are also represented. For 14 sources we found only generic radio source classification (radio, mm and sub-mm sources).}
\label{isto_emitters}
\end{figure}

We have used the Simbad database to check which stellar populations are identified in the field. We found a total of 2935 objects, 1774 of which are classified as stars (Fig.\,\ref{isto_emitters}). Among stars, there is a good spread in different classes of stellar objects thought to be radio emitters and a good representation of different spectral types.

Part of the proposed patch has already been surveyed in the mid-IR by SPITZER \citep{Benjamin_03, Carey2009} and by Herschel HI-GAL survey \citep{Molinari2010} and will be covered in the near future by CORNISH\footnote{CORNISH south is a high resolution survey of the Galactic Plane (PI: M. Hoare) being carried out with ATCA at 6 and $9\um{GHz}$ with a target rms noise of $0.18\um{mJy\,beam^{-1}}$} south, that will nicely complement the radio spectral information for the brightest detections.

\begin{figure}
\includegraphics[width=7cm]{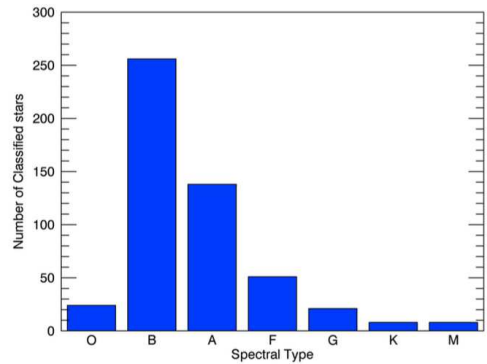}
\caption{Distribution of the subsample with Spectral Classification as reported in Simbad as function of spectral types. The larger number of B-type stars is a consequence of the presence of a part of the ScoOB1 association and of its nuclear cluster NGC 6231 in the field.}
\label{spectralTypes}
\end{figure}

\begin{figure*}
\includegraphics[width=7.cm]{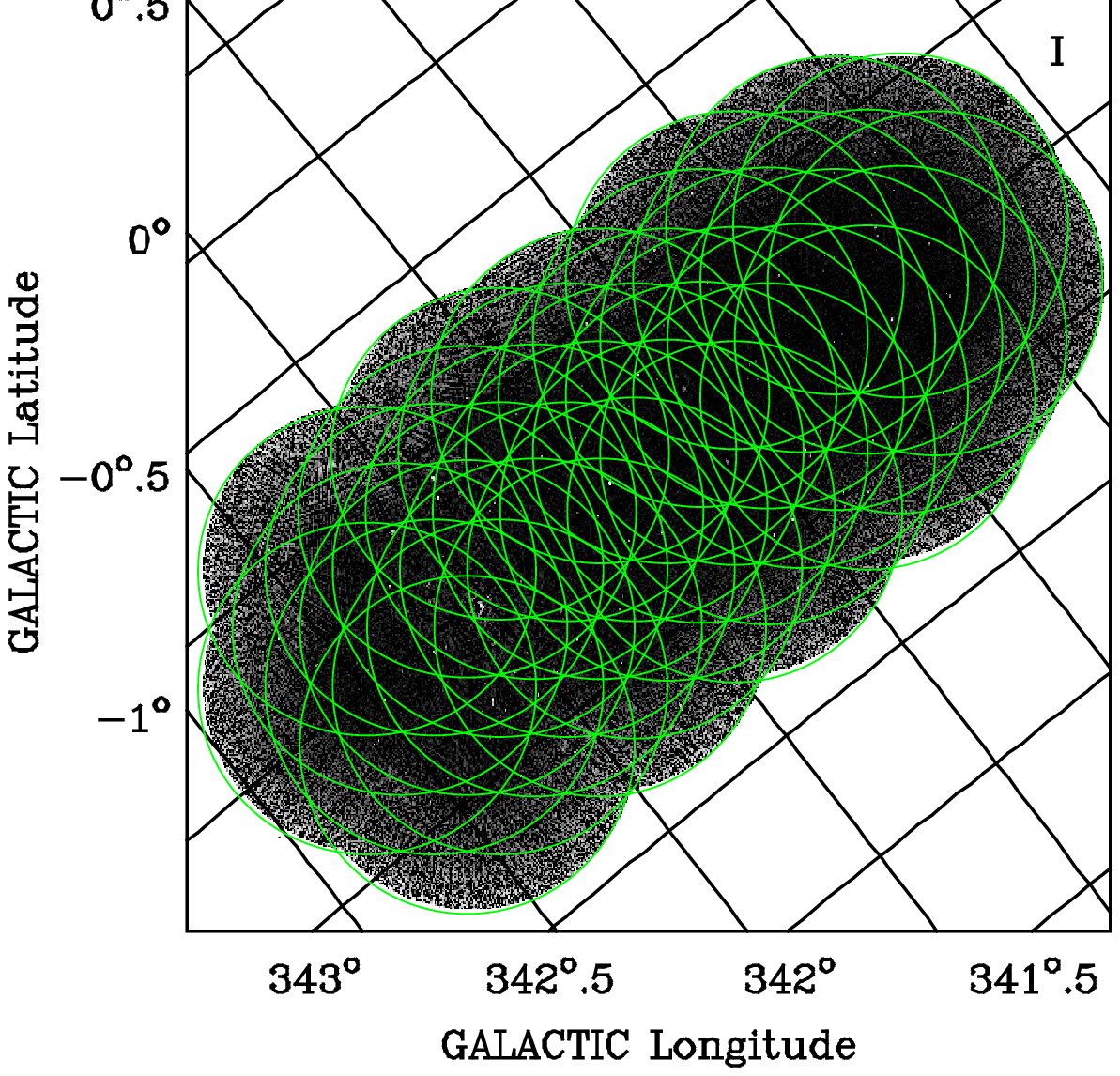}
\includegraphics[width=7.cm]{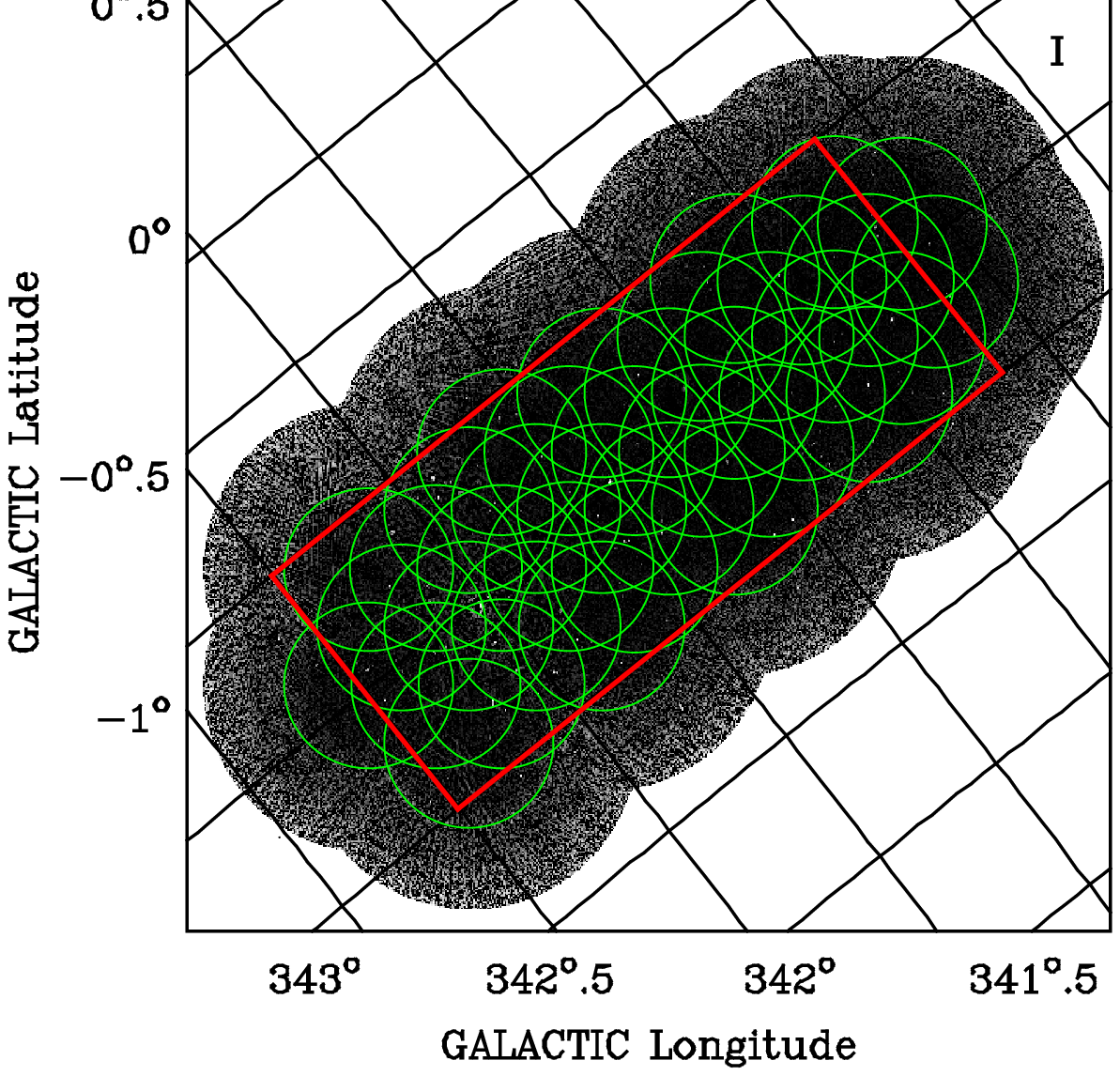}
\caption{Final map of the pilot field of Scorpio showing the ATCA field of view for each pointing at $1.470\um{GHz}$ (left panel) and $2.930\um{GHz}$ (right panel). The region chosen for source finding, including all the frequency sub-bands, is delimited by the red line. Grid coordinates are Galactic. Directions are RA (horizontal)-Dec (vertical).}
\label{sky_map}
\end{figure*}


\begin{table}
\caption{Log of Observations}
\begin{tabular}{lcc}
\hline
\hline

Date & Configuration & time on source \\
         &                           & (h)    \\
\hline
2011 April 21-24  & 6A & 38 \\
2012 June 3-11   & 6B  & 6 \\
\hline

\label{tab_observation}
\end{tabular}
\end{table}

\section{Observations and data reduction}
\subsection{The pilot experiment}
A pilot experiment for the SCORPIO project took place in 2011, from April 21 to 24  and in 2012, from June 3 to 11 (see Tab\,\ref{tab_observation}). In this experiment, only about one quarter of the selected field was observed, namely a $0.5\times2\um{deg}^2$ region centred  at $l=343.5^\circ$, $b=0.66^\circ$. The field was observed with ATCA, in mosaic mode, using a 8.8 arcmin spacing hexagonal grid, requiring a total of 38 pointings.

We cycled the array around these pointing centres, spending about $1\um{min}$ at each pointing centre. The secondary calibrator 1714-397 was visited every 20 min to calibrate the complex antenna gains. PKS B1934-638, which is the standard primary calibrator for ATCA observations, was  observed at the beginning of each observing run. Each  pointing was observed for $1.16\um{h}$ in total. The total observing time for all pointings was $58\um{h}$, including a 20 per cent overhead for flux, bandpass and phase calibration.

The observations were made with the new broadband backend system for the ATCA, the Compact Array Broadband Backend (CABB) \citep{wil2011}, with an effective observing band of $2\um{GHz}$ divided into 2048 1-MHz channels. All four Stokes parameters were measured.

RFI strongly corrupted our data, mostly between 1.2 and $1.4\um{GHz}$, requiring a large amount of flagging at those frequencies. Flagging and calibration were performed in MIRIAD, using the MIRFLAG task. We ended up flagging about 30-40 per cent of the  data. The usable range of frequencies goes from 1.350 to $3.100\um{GHz}$, corresponding to a frequency variation of about 80 per cent across the band. 

Calibration was performed by using the standard MIRIAD tasks. The well studied Gigahertz Peaked Spectrum radio galaxy PKS 1934-638 was used as bandpass and flux calibrator. A flux density of $12.31\um{Jy}$ at $2.1\um{GHz}$ \citep{reynolds1994} was used to derive the flux of the phase calibrator J1714-397 ($2.031\um{Jy}$ with an accuracy of $2\%$).

\begin{table}
\caption{Frequency sub-bands and primary beam radii at 5\% of the maximum amplitude}
\begin{tabular}{cccc}
\hline
\hline\
sub-band & central $\nu$ & $\Delta\nu$ &  R(5\%)  \\
               &   (GHz)        &      (GHz)    &     (arcmin)      \\
\hline\

 1 &     1.469 &      0.169 &         30.8    \\ 
 2 &     1.649 &      0.190 &         27.5    \\ 
 3 &     1.850 &      0.213 &         24.4    \\ 
 4 &     2.075 &      0.239 &         23.6    \\ 
 5 &     2.329 &      0.268 &         20.1    \\ 
 6 &     2.613 &      0.300 &         18.7    \\ 
 7 &     2.932 &      0.337 &         16.7    \\ 
\hline\
\label{channels}
\end{tabular}
\end{table}

\subsection{Map making}
Imaging was performed in MIRIAD and the deconvolution was done with the task MFCLEAN using the H\"ogbom algorithm \citep{hogbom}. All data for each pointing were combined into one file. Each pointing was then imaged separately prior to mosaicking. Five iterations of self-calibration were applied, the first three with phase calibration only and the last two with both phase and amplitude calibration, progressively increasing the number of CLEAN components used to model the sky emission. Multi-frequency CLEAN was used to model the spectral variation of the source emission using a two-term Taylor-Polynomial. The $uv$ data were weighted by adopting a robust weighting, with parameter 0.5. This permitted us to reach a better signal-to-noise ratio at the expense of a poorer angular resolution when compared to a uniform weighting scheme.

Next, the self-calibrated $uv$ data were divided into seven sub-bands, each of the same fractional bandwidth $\Delta\nu/\nu = 10\%$, imaged separately and adjusted to the reference position $\alpha=16\!:\!56\!:\!28.912$, $\delta=-42^\circ03'31.28''$. Splitting up the data into seven sub-bands serves to improve the accuracy of the primary beam correction. In Table \ref{channels}, the central frequency, bandwidth and primary beam size for each sub-band are listed; $R(5\%)$ is the radius of the primary beam out to a level of $5\%$ of the maximum, following the Gaussian primary beam model for ATCA telescopes \citep{wieringa_92}.

The $uv$ data were tapered in each sub-band differently to obtain nearly identical synthesized beams. The cell size was set to $1.5\um{arcsec}$, suitable for a good sampling of the synthesized beam. Pointings were CLEANed to a depth of $5\sigma$ and restored using a Gaussian beam of size 14.0 by 6.5 arcsec.

A mosaic for each sub-band was formed using the LINMOS task. Pointings were corrected for the primary beam and mosaicked together, weighting them by their respective rms noise values. The \texttt{bw} option was used to average the primary beam response over the sub-band (rather than use the primary beam response at the central frequency of the sub-band). Finally, the seven sub-band mosaics were combined using the IMCOMB task, assigning equal weights to the sub-bands.

In Fig.~\ref{sky_map}, the final map of the pilot experiment is shown. It is overlayed with the field of view (defined as the region out to $5\%$ of the primary beam maximum) of the ATCA antennas and centred on the pointing positions used in the mosaiced observations (see Tab.~\ref{channels}). In the left panel the field of view refers to the lowest frequency sub-band ($\nu\approx 1.470\um{GHz}$) and  in the right panel to the highest frequency sub-band ($\nu\approx 2.930\um{GHz}$).

\subsection{Mapping extended sources}
The standard CLEAN algorithm is an iterative procedure which assumes that the sky is a collection of point sources on an empty background. The algorithm looks for all these sources and creates a `CLEAN components' list, which is used to obtain a reliable representation of the sky by means of a convolution with an ideal PSF. This procedure models an extended source as a set of point sources. However the results may not be completely satisfactory.

Several algorithms have been proposed to properly model extended sources also. These methods assume that the sky is not composed of point sources only, but of sources of many different sizes and scales.

In particular, for our purpose, we used the multi-scale CLEAN algorithm \citep{Cornwell2008}, as implemented in CASA. With respect to the usual CLEAN procedure, we defined three different scales to model source extension: 0 to safely search for point sources (equivalent to a normal CLEAN), 5 pixels (about the beam dimension) and 15 pixels. A few examples of extended sources, imaged using the multi-scale CLEAN, are presented in Figs.\,\ref{DBS2003} and \ref{IRAS16566-4204}. Multi-scale CLEAN was found to be much more effective than standard CLEAN at removing the sidelobes around these sources.

\begin{figure}
\includegraphics[width=9cm]{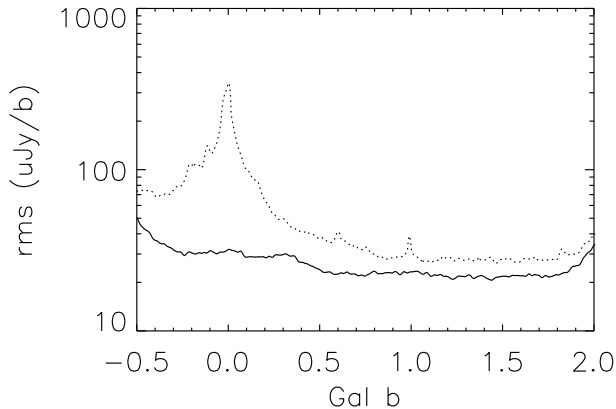}
\caption{The trend of rms (from the noise map) as a function of the Galactic latitude for both Stokes $I$ and $V$ (note the use of a logarithmic scale for the $y$ axis). For each pixel of the map the  rms has been averaged over $0.3\um{deg}$ in longitude. The increase in the rms close to the Galactic Plane  in Stokes I indicates the presence of diffuse emission.}
\label{rms}
\end{figure}

\section{Image Analysis and Source Catalogue}
\subsection{Source finding}
\label{sourcefinding}
Sources were detected through an analysis of the mosaic combining all the frequency sub-bands. A significant non-zero background is expected if there is large-scale, diffuse emission, which is the case in the SCORPIO field given its location on the Galactic Plane. Since our aim was to detect and characterize compact sources in the field, the background was measured and subtracted from the image before source extraction. The background was estimated for each pixel using the following procedure: because computing the background is fairly computationally expensive, it was only calculated for a limited number of pixels occuring at regular intervals in the $x$ and $y$ directions. For each of these pixels, the background was taken as the median pixel value inside a square centred on the pixel with a width of 10 times the synthesized beam, but a standard iterative technique was used where points were clipped if they deviated from the median by more than three times the rms of the deviation from the median in the previous iteration. The process was terminated when the rms of the deviation from the median from successive iterations differed by less than $10\%$. The background in the remaining pixels was estimated using a bicubic interpolation.

Having subtracted the background from the image, the next step involved determination of the noise level. For each pixel, an initial estimate of the noise was taken as the rms inside a box of size 10 times the synthesized beam, centred on the pixel. To prevent the noise estimate from being affected by real source emission, pixels outside the range $\pm3\sigma$ were removed and the rms was re-calculated. This process was repeated a number of times until the noise was found to decrease by no more than 10 per cent.

Sources were detected and characterized using a similar method to that employed by \citet{Franzen_2011} for the 10C survey. The noise maps were used to identify sources on the basis of their signal-to-noise ratio. Local maxima above $5\sigma$ were identified as sources. A peak position and flux density value were measured by interpolating between the pixels. A centroid position, integrated flux density and source area were also calculated by integrating contiguous pixels down to $2.5\sigma$, and sources were identified as overlapping if the integration area contained more than one source.

\subsection{Sensitivity}
There are a number of factors which affect the sensitivity of our SCORPIO mosaic: the effective bandwidth, system temperature, integration time and  other instrumental characteristics, as well as the ability to deconvolve extended sources, which in turn depends  on the {\it uv} plane sampling, in particular on the minimum baseline of the interferometer. The Galactic Plane is full of extended sources and diffuse synchrotron emission. In order to quantify the theoretical sensitivity limit, we made a $V$ polarization map of the whole field with the aim to compare the Stokes $I$ and $V$ rms maps.
   
In general, radio source emission presents a circular polarization close to 0\%. We can assume that no source is present in the $V$ map which therefore does not contain artifacts due to deconvolution errors and non-perfect $uv$ sampling. Stokes $I$ and $V$ are given by linear combinations of the cross correlations of the responses of the feeds, which are linearly polarized for ATCA ($X$ vertical and $Y$ horizontal polarizations):
\begin{equation}
2I=XX+YY
\end{equation}
and
\begin{equation}
2iV=XY-YX.
\end{equation}
Since all the cross products have the same statistics, $I$ and $V$ maps have the same statistics too, and therefore the noise level should be comparable. Fig.~\ref{rms} shows the rms for the $I$ and $V$ maps. At $b\approx0^\circ$ the $I$ map has a very high rms compared with the $V$ one. This is imputed to the presence of extended sources and diffuse emission in the Galactic Plane not correctly sampled in the $uv$-plane due to the lack of short baselines. This leads to fluctuations  which adds noise into the map. At the edge of the field ($b<-0.25^\circ$ and $b>1.8^\circ$) the noise increases due to the primary beam attenuation and the effective frequency  of the image decreases due to the frequency dependence of the primary beam (see Fig.~\ref{sky_map}). The rms of the $V$ map is in general lower, reaching about $20\mic{Jy\,beam^{-1}}$, while for the $I$ map it does not go below $30\mic{Jy\,beam^{-1}}$. In any case the high brightness of extended synchrotron emission at $|b|<0.4^\circ$ ($T_{B}\approx 10\,\um{K}$ at $16\,\um{cm}$) increases the rms of any polarization map.

\subsection{The compact components catalogue and Cross Identifications}
\label{sec:comp_comp_cat}
Radio components were extracted from the image as described in subsection~\ref{sourcefinding}. Sources were extracted only from a rectangular subregion of SCORPIO  with the best sensitivity. This is the region that includes the high frequency pointings and is defined by the rectangle between $l=343^\circ$ and $l=344^\circ$ and $b=-0.5^\circ$ and $b=1.833^\circ$, indicated in red in Fig.~\ref{sky_map}. Components corresponding to or affected by artifacts were removed. The resulting catalogue contains 614 radio components; Table \ref{tab_match} lists the first 25 components in the catalogue while the complete catalogue is available in the online version. Querying SIMBAD and NED (search radius of 6 arcsec) we found 34 matches. Some of these are classified as stars, PN or Pulsars, others  have only a generic classification (i.e. radio or IR source).

Among those 34  matches, we individuate 10 sources that are associated with objects that could be stars, 2 Planetary Nebulae (SCORPIO1\_614 and SCORPIO1\_325) and 2 Pulsars (SCORPIO1\_399 and SCORPIO1\_091). In the following list, more information on the possible stars are provided. The classification of these objects, as reported in the SIMBAD database, is indicated between square brackets.
\begin{itemize}
\item[][*] SCORPIO1\_534: generically classified in SIMBAD as star, no references are reported in the  SIMBAD database.
The source SCORPIO1\_534 is embedded within extended radio emission.
(see Section \ref{sec:IRAS_16561}).

\item[][IR*] SCORPIO1\_012: associated to  $\zeta^1$ Sco, classified by SIMBAD as Variable Star of irregular type.

\item[][WR*] SCORPIO1\_118: WR star, characterized by strong stellar wind already known as radio source  \citep{Bieging1982,abbott86,hogg89}. The wind radio emission has been detected at 1.3 mm ($11.6 \pm  2.3$ mJy) \citep{Leitheree1991}. The spectral index, estimated between  our SCORPIO measurement and the millimeter measurement, is consistent with radio emission from a stellar wind.

\item[][*] SCORPIO1\_311 (spectral type: WC7+O6V): well known WR star (HR 6265/HD 152270/WR 79) already known as radio source \citep{Bieging1982,abbott86,hogg89}.

\item[][iC*] SCORPIO1\_313 (Cl* NGC 6231 SBL 489):  
 classified as Star in Cluster, and was listed in the photometric study of the young open cluster NGC 6231 \citep{sung98}.

\item[][Y*?] SCORPIO1\_468 (G343.7018+00.086):  classified by SIMBAD as Young Stellar Object Candidate.

\item[][*] SCORPIO1\_243 (IRAS 16495$-$4140): classified as star in the SIMBAD database, no references reported.

\item[][Be*] SCORPIO1\_219  (V921Sco): well known Herbig Ae/Be star (128 references listed by SIMBAD). This star is surrounded by an extended envelop detected at the sub-mm wavelengths \citep{mannings95} and is also visible in the infra red.

\item[][*iC]  SCORPIO1\_406 (Cl* NGC 6231 SBL 759 ): classified as Star in Cluster, listed in the photometric study of the young open cluster NGC 6231 \citep{sung98}. The compact source SCORPIO1\_406 is embedded within more  extended radio emission. 

\item[][*iC] SCORPIO1\_350 (Cl* NGC 6231 BVF 77: (spectral type: B8.5V): classified as Star in Cluster and listed in the photometric study of the young open cluster NGC 6231 \citep{sung98}.

\end{itemize}

The first step in validating and  classifying the detected radio sources, with no matches in SIMBAD or NED, is to determine how many of these have counterparts  at other wavelengths. Hence, we cross-matched our radio sources with  major catalogues, from the optical to the radio frequency range, covering the SCORPIO region, namely: 
the Naval Observatory Merged Astrometric Dataset  (NOMAD; \citealp{zacharias_04}), 
the 2MASS all sky catalogue of point sources \citep{2mass_cat},
the Galactic Legacy Infrared Mid-Plane Survey Extraordinaire Source Catalogue (GIMPSE;  \citealp{glimpse_cat}),
the Wide-field Infrared Survey Explorer (WISE; \citealp{wise_cat}), 
the Midcourse Space Experiment infrared point source catalogue (MSX6C; \citealp{msx_cat}),
the AKARI mid-IR (AKARI/IRC; \citealp{akari_irc_cat}) all sky survey,
the Herschel infrared Galactic plane Survey catalogue (Hi-GAL; \citealp{Molinari2010}), 
the IRAS point source catalogue \citep{iras_cat},
the APEX Telescope Large Area Survey of the GALaxy at 345 GHz (ATLASGAL; \citealp{contreras_etal_2013}),
the 2nd Epoch Molonglo Galactic Plane Surveys (MGPS-2; \citealp{murphy_etal_2007}),
the catalogues of compact radio sources in the Galactic Plane (WBH2005; \citealp{White2005}) and
the radio survey of the Red MSX Source (RMS; \citealp{urquhart_etal_2007}).

\begin{table}
\caption{Number of matched sources } 
\begin{center}
\begin{tabular}{lc}
\hline
\hline
Catalog Name      &  Number of Matches  \\
\hline
NOMAD        &   320  \\
2MASS        &   301 \\
GLIMPSE      &   229  \\
WISE         &   116 \\
MSX6C        &   47   \\
AKARI        &   34   \\
Hi-Gal       &   148  \\
IRAS        &   117  \\
ATLASGAL        &   14  \\
MGPS-2         &   43  \\
WBH2005         &   18 \\
RMS         &   6  \\
\hline
\end{tabular}
\end{center}
\label{tab_match}
\end{table}

In performing such a  task the most critical information is position accuracy. The position errors in right ascension and declination for the radio sources were derived following equations 3, 4 and 5 in \citet{purcell13}, that take into account the effective signal-to-noise ratio for each point source and the interferometer beam size, assuming for the sky background a Gaussian correlated noise.

An error of 0.1 arcsec, corresponding to the position uncertainty of the phase calibrator \citep{Norris06}, was added in quadrature to this value, to obtain the final error on the radio position ($\sigma_\mathrm{radio}$). In the cross-correlation process, we set a search radius of $r=5\sigma_\mathrm{posi}$, where $\sigma_\mathrm{posi}=\sqrt{\sigma_\mathrm{radio}^2+\sigma_\mathrm{cat}^2}$ and $\sigma_\mathrm{cat}$ is the position error in the reference catalogues (and all other instances in paper).

The number  of sources in our catalogue that match the entries in the reference catalogues is summarized in Table~\ref{tab_match}. Among the 614 SCORPIO sources, 487 have a counterpart in at least one of the inspected catalogues. There are no SCORPIO sources concurrently  present in all inspected catalogues, however: 78 per cent of SCORPIO sources found in one of the IR catalogues (GLIMPSE, WISE, MSX) are also in NOMAD or 2MASS; 53 per cent of those found in one of the far-IR catalogues (AKARI, IRAS, Hi-GAL, ATLASGAL) are also in NOMAD or 2MASS; only 34 per cent of those found in one of the radio catalogues (RMS, WBH2005, MGPS-2) are also in NOMAD or 2MASS.


\noindent
The Compact Components Catalogue (Table \ref{tab_catalog}) is organized as follows:\\
Column (1) Component number. This is the internal definition of the component used within this paper. If two or more sources have  the same number (with different letters) they are components of a group, i.e. associated with  the same radio source, as returned by the adopted source finding method (see Section 4.1) and confirmed by a visual inspection.\\
Column (2) Source Name;\\
Column (3) Galactic longitude of the component;\\
Column (4) Galactic latitude of the component;\\
Columns (5) and (6) Right ascension and Declination (J2000.0) of the peak of the emission;\\
Columns (7) and (8) integrated flux density at $20\um{cm}$ ($S$) and its estimated uncertainty ($\Delta S$), defined as $\sqrt{\mathrm{rms}^2+(S\times\Delta W/W)^2+(aS)^2}$, in mJy \citep{Fom}; where $\mathrm{rms}$  is the local rms, $W$ and $\Delta W$ are the beam width and its associated error ($\Delta W=\mathrm{rms} \times W / P$, $P$ is the peak intensity) and $a$ is the relative error of the flux density of the calibrator. We assumed $a=0.03$;\\
Column (9) Area of the component in beam units; 1 means source not resolved;\\
Column (10) match with source in other catalogue. N: NOMAD, M: 2mass, G: GLIMPSE, W: WISE, X: MSX, A: AKARI, H: HIGAL, I: IRAS, a: Atlasgal, m: MGPS-2, w: WBH05, R: RMS.

\begin{table*}
\caption{First 25 entries of the Point Source Catalogue of the Pilot experiment. The complete catalogue is available in the online version.}
\begin{tabular}{lrrrrrccr}
\hline\hline
\multicolumn{1}{c}{ID} &  \multicolumn{1}{c}{$l$} & \multicolumn{1}{c}{$b$} & \multicolumn{1}{c}{RA} & \multicolumn{1}{c}{Dec}& \multicolumn{1}{c}{$S$} & \multicolumn{1}{c}{$\Delta S$} & \multicolumn{1}{c}{Area} & \multicolumn{1}{c}{Matching}\\
\multicolumn{1}{l}{  } &  \multicolumn{1}{c}{(deg)} & \multicolumn{1}{c}{(deg)} & \multicolumn{1}{c}{(J2000)} & \multicolumn{1}{c}{(J2000)}& \multicolumn{1}{c}{(mJy)} & \multicolumn{1}{c}{(mJy)} & \multicolumn{1}{c}{(beam)} & \multicolumn{1}{c}{}\\
\hline
SCORPIO1\_001 & 343.0025 & 1.7604 & 16:50:12.21& -41:48:56.2& 33.33 & 1.03 & 1.0 &-\\
SCORPIO1\_002 & 343.0051 & 0.2234 & 16:56:39.13& -42:47:08.2& 0.49 & 0.09 & 1.0 &-\\
SCORPIO1\_003 & 343.0134 & 0.6086 & 16:55:02.85& -42:32:15.5& 7.74 & 0.29 & 1.4 &NMGm\\
SCORPIO1\_004 & 343.0138 & 1.7208 & 16:50:24.33& -41:49:56.4& 1.60 & 0.17 & 1.0 &N\\
SCORPIO1\_005a& 343.0139 & 1.1508 & 16:52:46.37& -42:11:42.1& 12.80 & 0.50 & 2.0 &N\\
SCORPIO1\_006 & 343.0152 & 0.1166 & 16:57:08.49& -42:50:39.6& 1.27 & 0.16 & 1.7 &NMGW\\
SCORPIO1\_007 & 343.0157 & -0.1830 & 16:58:25.48& -43:01:49.4& 1.51 & 0.12 & 1.2 &NMG\\
SCORPIO1\_005b& 343.0186 & 1.1577 & 16:52:45.61& -42:11:13.0& 106.56 & 3.20 & 1.4 &-\\
SCORPIO1\_009 & 343.0201 & 1.5248 & 16:51:14.28& -41:57:09.1& 1.34 & 0.11 & 1.0 &-\\
SCORPIO1\_010 & 343.0203 & 0.3109 & 16:56:19.93& -42:43:08.3& 20.97 & 0.64 & 1.2 &Gmw\\
SCORPIO1\_011 & 343.0216 & 0.7407 & 16:54:31.13& -42:26:53.1& 1.12 & 0.10 & 1.3 &G\\
SCORPIO1\_012 & 343.0277 & 0.8702 & 16:53:59.71& -42:21:42.3& 1.17 & 0.09 & 1.2 &NMXAH\\
SCORPIO1\_013 & 343.0317 & 0.8579 & 16:54:03.63& -42:21:59.3& 0.44 & 0.08 & 1.0 &NMH\\
SCORPIO1\_014 & 343.0322 & 0.9655 & 16:53:36.65& -42:17:53.0& 1.21 & 0.12 & 1.3 &NH\\
SCORPIO1\_015 & 343.0356 & 0.2182 & 16:56:46.68& -42:45:53.7& 8.06 & 0.25 & 1.0 &N\\
SCORPIO1\_016 & 343.0394 & 1.5983 & 16:50:59.99& -41:53:26.8& 2.49 & 0.13 & 1.0 &I\\
SCORPIO1\_017 & 343.0415 & 1.2316 & 16:52:31.80& -42:07:20.7& 1.46 & 0.13 & 1.0 &NM\\
SCORPIO1\_018 & 343.0428 & 0.0266 & 16:57:37.15& -42:52:43.9& 5.29 & 0.22 & 1.0 &-\\
SCORPIO1\_019 & 343.0457 & 0.8848 & 16:53:59.73& -42:20:18.9& 71.55 & 2.15 & 1.0 &Hm\\
SCORPIO1\_020 & 343.0461 & 1.2752 & 16:52:21.85& -42:05:28.5& 4.71 & 0.25 & 2.2 &I\\
SCORPIO1\_021 & 343.0468 & 0.3010 & 16:56:27.87& -42:42:16.1& 0.47 & 0.06 & 1.0 &NMGW\\
SCORPIO1\_022a& 343.0489 & 0.9666 & 16:53:39.77& -42:17:04.1& 109.20 & 3.28 & 1.6 &Gm\\
SCORPIO1\_023 & 343.0501 & 1.6133 & 16:50:58.45& -41:52:22.7& 1.75 & 0.17 & 2.1 &NMWAI\\
SCORPIO1\_024 & 343.0502 & 0.4249 & 16:55:57.01& -42:37:27.4& 1.04 & 0.09 & 1.0 &NMGH\\
SCORPIO1\_022b& 343.0514 & 0.9634 & 16:53:41.09& -42:17:04.3& 23.35 & 0.72 & 1.6 &m\\
\hline
\end{tabular}
\label{tab_catalog}
\end{table*}

\section{Extended Sources: a few examples}
Our source extraction method is biased against extended sources. However, a visual inspection of the final map points out the presence of 17 of this kind of sources falling in the SCORPIO field. In particular, there are many roundish structures recalling what several authors named as "bubbles" \citep{church2006}. Bubbles are pervasive throughout the entire Galactic Plane and have been mainly discovered by means of the new extended survey conducted with Spitzer, namely GLIMPSE \citep{Benjamin_03} and MIPSGAL \citep{Carey2009} as they are quite often associated with extended dusty structures. 

\begin{figure}
\includegraphics[width=8.4cm]{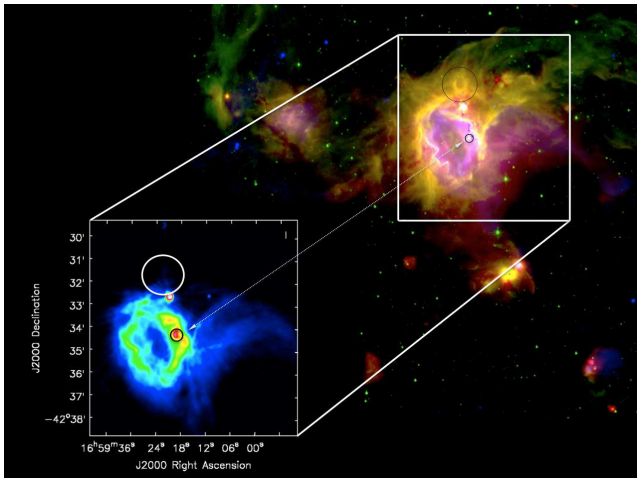}
\caption{Composite picture of the field centred on $[$DBS2003$]$~176. The sub-panel shows only the SCORPIO map, while in the background panel the mid-IR/FIR maps from Spitzer (IRAC, $8\mic{m}$, green) and Herschel (PACS, $70\mic{m}$, red) are superimposed on the SCORPIO map, in blue. The white arrow indicates the position of the the compact component SCORPIO1\_300. The white circle points out that there is no radio emission associated with S16 (see text).}
\label{DBS2003}
\end{figure}

Infrared bubbles are usually  associated with hot young stars in massive star formation regions and in some cases they are coincident with known H \textsc{ii} regions \citep{church2006}. However, there is  growing observational evidence that some bubbles can be related to massive evolved stars, such as LBV, WRs and SNRs \citep{Wachter2010, Gvaramadze2010} or with PNe \citep{Ingallinera2013}. Despite the apparent similar morphologies, there  are many differences between the emission characteristics which stand out clearly when different maps, obtained at different IR and radio bands, are compared. As an example, bubbles associated with  massive star formation regions have an extended $8\mic{m}$ emitting region embracing the more compact $24\mic{m}$ one, usually co-spatial with the ionized region, traced by the radio. This is interpreted as a signature of a PDR, traced by the PAHs emission at $8\mic{m}$ \citep{church2006}. In contrast, the emission characteristic of a Bubble associated with an evolved star has a more complex structure. The dusty envelope is usually brighter and more extended at $24\mic{m}$ \citep{Wachter2010} but the ionized  region can be  well contained or co-spatial or even more extended with respect to the  $24\mic{m}$ emission  \citep{Chu2009}. This is interpreted as a result of the very complex dust distribution in the circumstellar envelope as consequence of different mass-loss episodes that can occur during the evolution of the central star.

In the following we will provide a few  examples of  extended sources detected in the SCORPIO pilot. Some of these  are associated with  already known Galactic sources, while others  are unclassified. Our work provides the first high resolution radio continuum maps for most of these sources, revealing the details of the ionized gas.

In presenting our results we  also consider  the dusty environment associated with  the radio sources and display the radio maps together with maps of the regions provided by GLIMPSE and HIGAL \citep{Molinari2010}, that trace the warm and cooler dust distribution in the surrounding regions.

\subsection{$[$DBS2003$]$~176}
The stellar cluster $[$DBS2003$]$~176 is associated with a multiple bubble (S16 and S17) detected in the GLIMPSE survey \citep{church2006}. The H \textsc{ii} region associated with the bubble S17 has been  detected in the Parkes-MIT-NRAO (PMN) survey \citep{Wright1994} at $4.5\um{GHz}$, which has a beam size of $4.1\um{arcmin}$. Radio emission at $20\um{cm}$ from an extended non-Gaussian source has also been reported by \cite{Zoo_1990}.

Our SCORPIO map reveals, for the first time, the extended radio emission at $20\um{cm}$, with unprecedented  detail, thanks to an angular resolution comparable to mid-IR/FIR  images. In Fig \ref{DBS2003}, the radio emission (SCORPIO) is  pictured in blue, the Herschel (Hi-GAL) map at $70\mic{m}$ in red, and the Spitzer (Glimpse) map at $8\mic{m}$ in green. The extended radio emission is well contained inside the dust emission, traced by the IR. In the false color image the associated stellar cluster is recognizable inside bubble S17. This stellar cluster is spatially resolved in the mid-IR ($8\mic{m}$) in many stellar components.

The two bubbles, S16 and S17, are clearly visible in the IR maps, while there is no extended radio emission associated with bubble S16 (sub-panel). Two point-like radio sources (SCORPIO1\_320 and SCORPIO1\_300) have been retrieved by our extraction procedure. The position of SCORPIO1\_300 in the composite RGB map is indicated by the arrow, see Fig.\ref{DBS2003}.
The brighter radio point  source (SCORPIO1\_320), clearly visible within the two IR bubbles S16 and S17, is an already known radio source \citep{Zoo_1990}  and is present in The Two Micron All Sky Survey (2MASS) catalogue of extended sources \citep{Skru_2006}. A flux density measurement at $5\um{GHz}$ (176.9 mJy) has also been reported for SCORPIO1\_320 and the source has been tentatively classified as a Massive Young Stellar Object (MYSO) candidate \citep{urquhart_etal_2007}. Source SCORPIO1\_300 is located within bubble S17 and has already been detected at  radio wavelengths by \citet{Zoo_1990}. It is also associated with a GLIMPSE source. From  Fig.\ref{DBS2003}, SCORPIO1\_300 seems to be located in the top region of a dust pillar, well visible at the IR wavelengths, indicating possible triggered star formation by the expanding H \textsc{ii} region. Further investigation of that source is necessary to identify it as a YSO.

\subsection{IRAS 16566-4204}
\begin{figure}
\includegraphics[width=8.4cm]{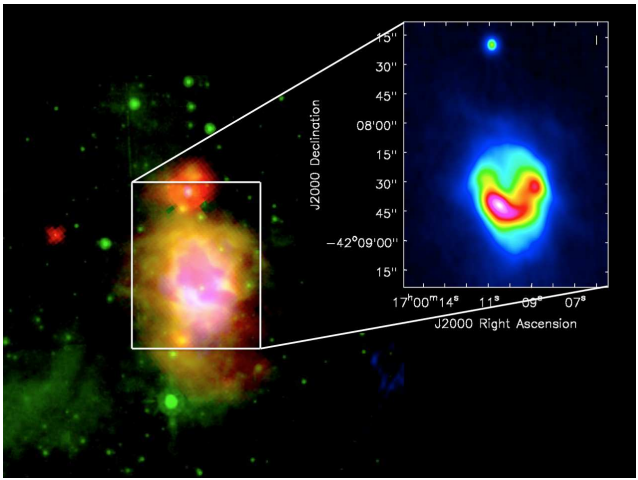}
\caption{ Composite picture of the field around IRAS 16566-4204. In the field two radio sources are visible. The point-like source is associated with  SCORPIO1\_600, while the more extended source is associated with SCORPIO1\_592a and SCORPIO1\_592b. The mid-IR/FIR maps from Spitzer (IRAC, $8\mic{m}$, green) and Herschel (PACS, $70\mic{m}$, red) are superimposed to the SCORPIO map (blue) with colour coding as indicated.}
\label{IRAS16566-4204}
\end{figure}

In Fig.\,\ref{IRAS16566-4204}, the field around the position of IRAS 16566-4204 is shown. Two radio sources are clearly visible, one very compact (SCORPIO1\_600) and the other more extended. Their positions are coincident with those of IRAS 16566-4204. IRAS 16566-4204 has been previously classified as a  UCHII on the basis of IRAS colour-colour criteria \citep{Wood1989} and of the association with both radio continuum and maser emission \citep{Walsh1997}. However, higher resolution ATCA observations revealed a compact radio source and a masing methanol (6.668 GHz) source with a position offset with respect  to the IRAS pointing center \citep{Walsh1998}. Such a picture together with the SCORPIO map is consistent with the UCHII  being associated with  SCORPIO1\_600 and IRAS 16566-4204 associated with the extended radio source whose classification is not clear. However, the radio emission is well contained inside the more extended $8\mic{m}$ emission (see Fig\,\ref{IRAS16566-4204}), providing us with a morphological hint to classify IRAS 16566-4204 as a massive star formation region. At the position of IRAS 16566-4204 a $101\um{mJy}$ source at $4.85\um{GHz}$ is reported  in the  Parkes-MIT-NRAO (PMN) survey.

\begin{figure}
\includegraphics[width=8.4cm]{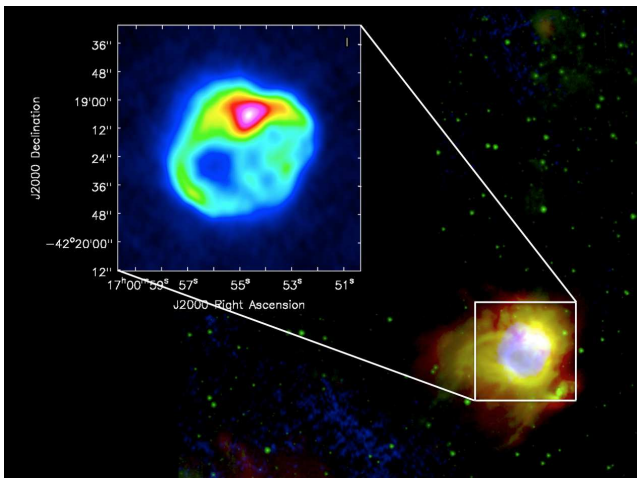}
\caption{Composite picture of the field centred at the position of  IRAS16573-4214.
The mid-IR/FIR maps from Spitzer (IRAC, $8\mic{m}$, green) and Herschel (PACS, $70\mic{m}$, red) are superimposed on the SCORPIO map (blue) with colour coding as indicated.}
\label{IRAS16573}
\end{figure}

\subsection{IRAS 16573-4214}
The source has been classified as a bona fide high-mass protostellar object  on the basis of its IR colours. The source was observed at $6.7\um{GHz}$ by \citet{MacLeod1998} to search for methanol maser but this was not detected. The source was instead detected in the molecular lines of CS and C$^{17}$O by \citet{Fontani2005}, implying its association with  molecular clouds. \citet{San2013} recently observed the field with ATCA to search for  H$_{2}$O maser emission. While radio continuum emission at 18 and $22.8\um{GHz}$ was detected at the IRAS position, an H$_{2}$O maser was detected $\approx 20''$ to the edge of the continuum emission. Continuum emission at $1.2\um{mm}$ coincident with the H$_{2}$O maser was detected with the SEST telescope \citep{Beltran2006}.
  
The field around  the IRAS source IRAS16573-4214 is shown in Fig. \ref{IRAS16573}, as a superimposition of the radio image on  mid-IR/FIR  maps from Spitzer (IRAC, $8\mic{m}$) and Herschel (PACS, $70\mic{m}$). The SCORPIO map, obtained at 2.1 GHz with an improved spatial resolution with respect to those by \cite{San2013}, reveals much more detail of the radio source morphology and in particular a disk-like shape, well embedded  in a dusty environment as traced by the $70\mic{m}$ emission. The H$_{2}$O maser emission is not associated with  IRAS16573-4214 and we do not detect any radio continuum  emission at the position reported by \citet{San2013}, at our local sensitivity  threshold ($5\,\sigma$) of $500\mic{Jy}$.

\subsection{IRAS 16561-4207}
\label{sec:IRAS_16561}
\begin{figure}
\includegraphics[width=8.4cm]{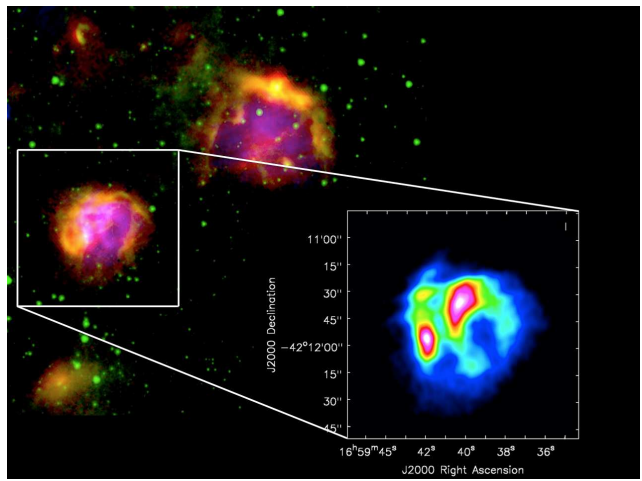}
\caption{Composite picture of the field centred at the position of IRAS 16561-4207. The mid-IR/FIR maps from Spitzer (IRAC, $8\mic{m}$) and Herschel (PACS, $70\mic{m}$) are superimposed on the SCORPIO map with colour coding as indicated.}
\label{Bubble1}
\end{figure}

In Fig.\ref{Bubble1}, the SCORPIO region associated with IRAS 16561-4207 is shown superimposed on mid-IR/FIR maps from Spitzer (IRAC, $8\mic{m}$) and Herschel (PACS, $70\mic{m}$), with colour coding as indicated. The brightest source on the left is SCORPIO1\_534, classified as a generic {\em star} in the SIMBAD database with no references associated. The source is imbedded in  extended radio and IR emission. Both $8\mic{m}$ and $70\mic{m}$ share the same morphology as well as the same extension, while the radio emission appears to be well contained inside the dusty cocoon. \citet{Walsh2014} reported the detection of a $\mathrm{H}_{2}\mathrm{O}$ maser spot at $\alpha=16\!:\!59\!:\!34.3$, $\delta=-42^\circ11'59.8''$. It is not clear if this maser spot is associated or not to the source, being localized in the external border of the IR emission.

\subsection{IRAS 16520-4146}
\begin{figure}
\includegraphics[width=8.4cm]{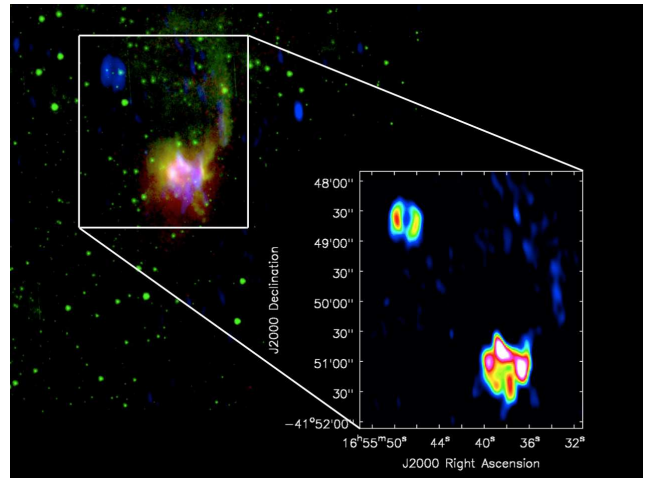}
\caption{ Composite picture of the field around the position of IRAS 16520-4146. The mid-IR/FIR  maps from Spitzer (IRAC, $8\mic{m}$, green) and Herschel (PACS, $70\mic{m}$, red) are superimposed on the SCORPIO map (blue) with colour coding as indicated. In the insert in the low right corner only the SCORPIO map (blue) is shown.}
\label{TwoBubbles}
\end{figure}

In Fig.\ref{TwoBubbles}, the SCORPIO  field around IRAS 16520-4146 is shown superimposed on mid-IR/FIR maps from Spitzer (IRAC, $8\mic{m}$) and Herschel (PACS, $70\mic{m}$), with colour coding as indicated. To better appreciate the radio morphology of the sources detected in the SCORPIO field,  only  the radio data are  shown in the insert in the bottom right corner of the same figure.
 
In the SCORPIO map two extended components are evident: the one in the south-west  is associated with IRAS 16520-4146, while the source in the north-east is not classified in the SIMBAD database but is reported in the Mizuno catalog of mid-IR Galactic bubbles.  Moreover, while IRAS 16520-446 has a mid-IR/FIR  counterpart, this is not the case  for the other source in the field, that however shows a clear roundish morphology at $24\mic{m}$ \citep{Mizuno2010}.  Interestingly, a hard X-ray transient was detected by INTEGRAL (IGR J16558-4150) at $\alpha=253.95^\circ$, $\delta=-41.83^\circ$, with 3-arcmin accuracy \citep{Soldi2007}. No radio counterpart is visible at the coordinates of the INTEGRAL source. However, due to the low accuracy of the INTEGRAL position we cannot exclude that the unclassified radio source, detected to the north-east of IRAS 16520-446, is related to the X-ray source. Further investigation is necessary.
 
\section{Some preliminary conclusions}
The aim of this  paper is to provide an overview of the SCORPIO project and to present the results of the pilot observations. Some conclusions can   however be drawn on the basis of these preliminary results. In spite of the  large number of objects located in the selected field and  classified as stars, according to SIMBAD, we found very few matches that definitively associate SCORPIO radio components with stellar objects. Among the  matches with the SIMBAD database, 10 sources are classified as stars. All of them appear related to thermal emission  and associated with an early-type star (see Section \ref{sec:comp_comp_cat}). However, this could be an effect of the higher number of stars with spectral classification `B-type' in the selected field as indicated by the distribution of spectral types shown in Fig.\,\ref{spectralTypes}.

The mere presence in the field of stars belonging to a class thought to be radio emitting is a necessary but not sufficient condition to detect them, as a series of requirements needs to be satisfied: a sufficient mass-loss rate and ionizing UV field, for thermal emitters, and a sufficient magnetic field strength and energetic particle density, for non-thermal ones. Distances also play  their role and variability is expected in most of the non-thermal radio emitters.

All the above considerations translate into a real difficulty to quantify the number of stellar objects we expect to detect in a specific patch of the sky.

We have  detection rates going from 20 per cent for OB stars \citep{Bieging1989} and 25 per cent for CP stars \citep{Trigilio2004} to 30-40 per cent for the active binary systems \citep{Umana1998}. However, all information  collected so far comes from small samples biased towards some kind of peculiarities observed in other spectral regimes and the quoted detection rates are to be considered as upper limits and are probably overestimated.

As  indicated in section 4.4, 487 of the 614 SCORPIO sources have a counterpart in at least one of the inspected catalogues. In particular, for our purposes, it is important to analyse how many SCORPIO sources have matches in stellar catalogues. The  NOMAD  (Naval Observatory Merged Astrometric Dataset) catalogue is a major stellar catalog which contains data from the Hipparcos, Tycho-2, UCAC-2 and USNO-B1 catalogues and is supplemented by photometric information from the 2MASS final release point source catalogue. As indicated in Table \,\ref{tab_match}, most of the SCORPIO sources found in NOMAD do not have an optical counterpart (301/320 NOMAD matches are actually from 2MASS).

This may indicate that a significant fraction of the  SCORPIO sources represents  a reddened stellar/galactic population. It is possible that this reddened population has some contaminants, namely some background galaxies. In principle, we may use the WISE colours to separate extragalactic versus Galactic sources, as those have revealed to be very effective in extragalactic surveys  \citep{Jarrett_2011}. Stars are characterized by WISE colours clustering around zero magnitude while different classes of extragalactic objects are clearly separated from them because of their redder colours. However, there are some Galactic objects, with extended dusty envelopes (AGB stars, YSOs and PNe) that have WISE colours similar to those of the different  extragalactic populations and they occupy the same regions in the colour-colour diagram \citep{Nikutta_2014}. Therefore, it appears that, in our case,   the WISE colour-colour diagram is of limited use for pointing out possible extragalactic contaminants and further ancillary observations are necessary.

Our observations, with  unprecedented sensitivity and angular resolution at $2.1\um{GHz}$, have allowed, for the first time, to point out radio emission associated to some extended sources, many of which are classified as Galactic bubbles. For some of them the SCORPIO map is the first radio detection, while for the sources already known to be associated with a radio source, the SCORPIO map constitutes an impressive improvement with respect previous observations.
Furthermore, comparison with existing mid-IR/FIR images has helped to classify some of the objects.

\section*{Acknowledgments}
This research is supported by ASI contract I/038/08/0 ``HI-GAL''.
This research made use of Montage, funded by the National Aeronautics and Space Administration's Earth Science Technology Office, Computation Technologies Project, under Cooperative Agreement Number NCC5-626 between NASA and the California Institute of Technology. Montage is maintained by the NASA/IPAC Infrared Science Archive.

\end{document}